\documentclass[aps,prb,twocolumn,longbibliography,groupedaddress,showpacs,showkeys,citeautoscript,amsfonts,amsmath,amssymb,flushbottom,floatfix]{revtex4-1}

\usepackage[T1]{fontenc}
\usepackage[latin1]{inputenc}
\usepackage{graphicx}
\usepackage{nicefrac}
\usepackage{amsmath}
\usepackage{mathtools}
\usepackage{wasysym}
\usepackage{MnSymbol}
\usepackage{url}

\begin{document}
\newcommand{\ds}{\displaystyle}
\newcommand{\D}{\mathrm{d}}
\newcommand{\I}{\mathrm{i}}
\newcommand{\EXP}[1]{\mathrm{e}^{#1}}

\newcommand{\ptt}{\tilde{\varphi}}
\newcommand{\abs}[1]{\left\lvert #1 \right\rvert}

\renewcommand{\vec}[1]{\boldsymbol{#1}}
\newcommand{\evecr}{\vec{e}_r}

\newcommand{\abl}[3][\empty]{\frac{\D^{#1}#2}{\D{#3}^{#1}}} 
\newcommand{\pabl}[3][\empty]{\frac{\partial^{#1}#2}{\partial{#3}^{#1}}} 

\newcommand{\gR}{\bar{g}}

\title{Dust ball physics and the Schwarzschild metric}

\author{Klaus Kassner}
\affiliation{Institut für Theoretische Physik, \\
  Otto-von-Guericke-Universität Magdeburg, Germany }

\begin{abstract}
  A physics-first derivation of the Schwarzschild metric is given.
  Gravitation is described in terms of the effects of tidal forces (or of
 spacetime curvature) on the volume of a small ball of test
  particles (a dust ball), freely falling after all particles were at
  rest with respect to each other initially. The possibility to
  express Einstein's equation this way and some of its ramifications
  have been enjoyably discussed by Baez and Bunn [Am.~J.~Phys.~\textbf{73}, 644
  (2005)].  Since the formulation avoids the use of tensors, neither
  advanced tensor calculus nor sophisticated differential geometry are
  needed in the calculation. The derivation is not lengthy and it has
  visual appeal, so it may be useful in teaching.
\end{abstract}
\date{December 9, 2016}


\maketitle

\allowdisplaybreaks

\section{Introduction}
\label{sec:intro}

With the increasing scope of its applications, general relativity (GR)
has become a subject that is taught more and more frequently already
in (upper-level) undergraduate courses. A \emph{physics-first}
approach has been developed and
advocated,\cite{hartle03,chen05,hartle06} in which the physical
consequences of interesting metrics are explored before Einstein's
field equations.\footnote{I will use the terms \emph{Einstein's field
    equations}, \emph{Einstein's equation}, and just \emph{field
    equations} interchangeably. They all mean the same thing. }
Similar goals are pursued in the so-called
\emph{intertwined$\,\,+\,$active-learning}
approach.\cite{christensen12,moore13} The Schwarzschild metric, in
particular, is a useful tool in such a program, as it permits the
quantitative discussion of the four classical tests of GR.

While this strategy is viable as a means
leading to increased interest in, and deeper understanding of, general
re\-lativistic phenomena, it has the disadvantage that the metric will
appear out of nowhere and its justification must wait until a lot of
mathematics has been learned. It would then seem desirable to be able
to obtain the met\-ric from simple arguments avoiding the full glory and
difficulty of tensor calculus and differential geometry.

These arguments must somehow replace the field equations in a
derivation of the metric, as it is well-known that a metric describing
curved spacetime cannot be derived without some ingredient going
beyond the Einstein equi\-valence principle (EEP) combined with the
Newtonian limit (NL). Detailed discussions why this is true for the
Schwarzschild metric have been given in
Refs.~\onlinecite{schild60,gruber88}.

In a recent article,\cite{kassner15} I have shown how the single
postu\-late that the gravitational field is source-free in vacuum
together with the experimental result on Mercury's peri\-helion
precession may be used to derive an approximate form of the
Schwarzschild metric capturing the first-order effects of all the
classical tests. The calculation is not difficult. However, deriving 
the perihelion shift with a slightly more general metric than the
Schwarzschild one proves to be a bit lengthy. Some may 
consider it preferable to do the perihelion shift calculation
only with the final form of the metric, which has fewer terms
than the expansion used in Ref.~\onlinecite{kassner15}.

Using \emph{two} postulates, it is possible to derive the exact form
of the Schwarzschild metric with little effort.\cite{kassner16} Yet,
the approach discussed in Ref.~\onlinecite{kassner16} requires
a profound understanding of wave phenomena and  is the outcome of
a research project rather than a classroom method. It might be used in a
class of exceptionally gifted students, demanding a flexibility of
thinking that even some teachers in the field of relativity may not
muster.  This suggests to look for yet another approach that is both
less computationally demanding than the one from
Ref.~\onlinecite{kassner15} and not as radically innovative as the one
from Ref.~\onlinecite{kassner16}.

As it turns out, there is an additional way. It produces the
Schwarzschild metric on the basis of the physical contents of the
field equations but neither requires advanced tensor calculus nor
mastery of differential geometry. The author became aware of this
possibility through the beautiful paper \emph{The meaning of
  Einstein's equation} by J. C.  Baez and E. F. Bunn.\cite{baez05}
They explain the physics of the field equations in terms of the volume
dynamics of a dust ball falling freely in the gravitational field.
The physical law embodied in the field equations may be described in a
single sentence that deserves to be cited:\cite{baez05} ``\emph{Given
  a small ball of freely falling test particles initially at rest with
  respect to each other, the rate at which it begins to shrink is
  proportional to its volume times: the energy density at the center
  of the ball, plus the pressure in the $x$ direction at that point,
  plus the pressure in the $y$ direction, plus the pressure in the $z$
  direction.}'' To distinguish it from the field equations proper, I
will call this the dust ball (DB) law (of gravitation) in the
following.\footnote{Note that in general, the condition of the dust
  particles in the ball initially being at rest with respect to each other
  makes sense only for a sufficiently small ball. General relativity
  does not allow us to assign a physical meaning to relative
  velocities, hence a relative state of rest, for objects that are not
  very close to each other.
}

In the case of vacuum, there is no energy density nor pressure, so we
may state the essence of the vacuum field equations in a tensor-free
formula
\begin{align}
\left.\frac{\ddot{V}}{V}\right\rvert_{\tau=0} = 0 \>,
\label{eq:vol_const}
\end{align}
where a dot signifies a derivative with respect to~the proper time $\tau$ of
the center particle of the ball and $V$ is the ball's volume as measured by
the center particle. Let us term this the \emph{DB vacuum equation}.

Naturally, particles are assumed to be so small that the attraction
between them due to their own gravitational field is negligible.  Baez
and Bunn warn that the DB law is not the formulation of Einstein's
equation that is most easily applied in various general relativistic
settings. Often the field equations in tensorial form are easier to
use and better suited to doing calculations. But it is certainly true
that Eq.~\eqref{eq:vol_const} is a physics-first formulation
expressing the essence of the vacuum field equations (with the fine
print added by Ref.~\onlinecite{baez05} that it is fully equivalent to
the field equations only, if the equation is required to hold for
small balls of arbitrary initial velocities). To state and apply the
law no advanced mathematical tools such as covariant derivatives,
higher-rank differential forms or the curvature tensor are needed.

Clearly, Eq.~\eqref{eq:vol_const} could be used in a GR course at an
early stage, with the promise that its equivalence to the (vacuum)
field equations would be proven later. This proof will not be given
here, it is contained in Ref.~\onlinecite{baez05}. The purpose of this
paper, then, is to use the DB vacuum equation in addition to the EEP
[incorporating special relativity (SR)] and the NL 
to derive the metric outside a spherically symmetric
mass distribution. We shall see that this is not entirely trivial but
it is an exercise worth doing and it might be of use in class\-room,
yielding a true physics-first derivation of the Schwarzschild metric.

The remainder of this paper is organized as follows.
Section~\ref{sec:schwarzschild} introduces the general form of a
stationary spherically symmetric metric compatible with the choice of
the circumferential radius as a radial coordinate and with time
orthogonality. Then, the strategy for the calculation of the rate of
volume change \eqref{eq:vol_const} based on the geodesic equations is
discussed. A subtlety in the evaluation of the comoving volume is
exhibited in a practice calculation.  Finally, the rates of volume
change of dust balls falling on radial and circular trajectories,
respectively, are computed. These two scenarios are visualized in
Figs.~\ref{fig:radial_fall} and \ref{fig:circular_fall}. Setting
${\ddot{V}}/{V}\rvert_{\tau=0}$ equal to zero constrains the metric in
the form of a differential equation for the coefficient functions, in
either of the configurations, and the two differential equations turn out
to be independent. They are solved in Sec.~\ref{sec:solution_metric},
which completes the derivation.  Section~\ref{sec:conclusions}
concludes by discussing two philosophies of presenting the method.

\section{The metric of a spherically symmetric mass distribution}
\label{sec:schwarzschild}

The sought-for metric must be spherically symmetric and stationary.
After fixing coordinates, it will depend on two independent radial
functions only.  We may write the line element in the form
\begin{align}
\D s^2 &= -f(r) \,c^2 \D t^2 + h(r) \, \D r^2 
+ r^2 \left(\D \vartheta^2 + \sin^2 \vartheta \,\D \varphi^2\right)\>,
\label{eq:spherical_static}
\end{align}
where $t$ is a temporal, $r$ a radial coordinate and $\vartheta$ and
$\varphi$ are standard angular coordinates.  While the most general
form of a spherically symmetric stationary metric contains four
functions of the radial coordinate,\cite{kassner16} two of these can
be eliminated by a choice of coordinates. The prefactor of the angular
coordinate term could be some arbitrary positive function. Choosing as
radial coordinate the circumference of a circle about the symmetry
center, divided by $2\pi$, we fix the prefactor to be $r^2$. Moreover, a
mixed term $\propto \D t\D r$ is possible in principle but can be
removed by a synchronization transformation of the time
coordinate.\cite{kassner16}

The EEP implies the geodesic equations for particles falling freely in
the metric, as has been discussed in some detail in
Ref.~\onlinecite{kassner15}.  These equations may be obtained from the
Lagrangian\footnote{In standard relativistic Lagrangian mechanics, the
  action of a free particle between two events is, up to a prefactor,
  the integral of the line element $\int_1^2 \D s$. The Lagrangian is
  then proportional to $\D s/\D t$. The Lagrangian
  \eqref{eq:spherical_lagrangian} is, apart from a prefactor, the
  square of this, but with the arbitrary time coordinate $t$ replaced
  by the proper time $\tau$. It can be shown that, if an affine
  parameter $\tau$, provided by the proper time for massive particles,
  is chosen as time coordinate in the action integral, extremalisation
  of $\int_1^2 (\D s/\D \tau)^2 \D\tau$ yields the same equations of
  motion as extremalisation of $\int_1^2 \D s/\D t\,\D t$. Therefore,
  our Lagrangian produces the correct equations of motion -- and it is
  easier to use than the standard Lagrangian, avoiding the appearance
  of certain square roots.  The factor $\frac{1}{2}$ has been introduced
  for convenience, to cancel out some factors of 2, appearing in
  taking derivatives.  Finally, that $L$ is constant  is of course due
  to the fact that $L=\frac12 (\D s/\D\tau)^2$ and that $\D s^2 =
  -c^2\D \tau^2$ for massive particles.}
\begin{align}
  L &= \frac12\left[-f(r) \,c^2 \dot t^2 + h(r) \, \dot r^2 + r^2
    \left(\dot \vartheta^2 + \sin^2 \vartheta \,\dot
      \varphi^2\right)\right] = -\frac{c^2}{2} \>,
\label{eq:spherical_lagrangian}
\end{align}
where a dot denotes, as before, a derivative with respect to proper time. 
They read (with a prime denoting a deri\-vative with respect to the argument)
\begin{align}
&\abl{}{\tau} \left[f(r) \dot t\right] = 0\>,
\label{eq:tmotion}
\\
&\abl{}{\tau} \left[h(r) \dot r\right]+\frac12 f'(r) c^2 \dot t^2 
- \frac12 h'(r) \dot r^2-r \dot\vartheta^2\nonumber\\&\hspace*{2cm}
 - r\sin^2 \vartheta \dot \varphi^2 = 0\>,
\label{eq:rmotion}
\\
&\abl{}{\tau} \left[r^2 \dot\vartheta \right] - r^2 \sin\vartheta \cos\vartheta \dot \varphi^2 = 0\>,
\label{eq:thetmotion}
\\
&\abl{}{\tau} \left[r^2 \sin^2\vartheta \dot\varphi \right] = 0\>.
\label{eq:phimotion}
\end{align}
Equations \eqref{eq:spherical_lagrangian} through \eqref{eq:phimotion}
constitute five equations for four variables, but only four of them
are independent.  It is often useful to replace one of the
four equations of motion by \eqref{eq:spherical_lagrangian}, a first integral, therefore
all five equations have been written out here.

We may assume our initial dust ball to be spherical. During a
sufficiently short time interval, it will then deform into an
ellipsoid under tidal forces. By choosing a sufficiently symmetric
initial state, we are able to predict the directions of the semiaxes.
The volume of an ellipsoid with semiaxes $\ell_a(\tau)$,
$\ell_b(\tau)$, and $\ell_c(\tau)$ is given by $V(\tau)=\frac{4\pi}{3}
\ell_a(\tau) \ell_b(\tau) \ell_c(\tau)$, and using 
that the dust particles are at rest with respect to each other at
$\tau=0$, we find for the rate of volume change described by
Eq.~\eqref{eq:vol_const}
\begin{align}
  \left.\frac{\ddot{V}}{V}\right\rvert_{\tau=0} =
  \left.\frac{\ddot{\ell_a}}{\ell_a} \right\rvert_{\tau=0}+
  \left.\frac{\ddot{\ell_b}}{\ell_b}\right\rvert_{\tau=0} +
  \left.\frac{\ddot{\ell_c}}{\ell_c}\right\rvert_{\tau=0}\>.
\label{eq:vol_change}
\end{align}
(The denominators on the r.h.s. are all equal, if our initial
ellipsoid is a sphere.) What we have to do, essentially, is to
calculate, for a short time interval, the trajectories of the center
particle and of three particles at the ends of the three semiaxes, in
order to obtain the relative rate of volume change, and to set this equal to
zero at $\tau=0$. 

\begin{figure}[t]
\includegraphics[width=7.2cm]{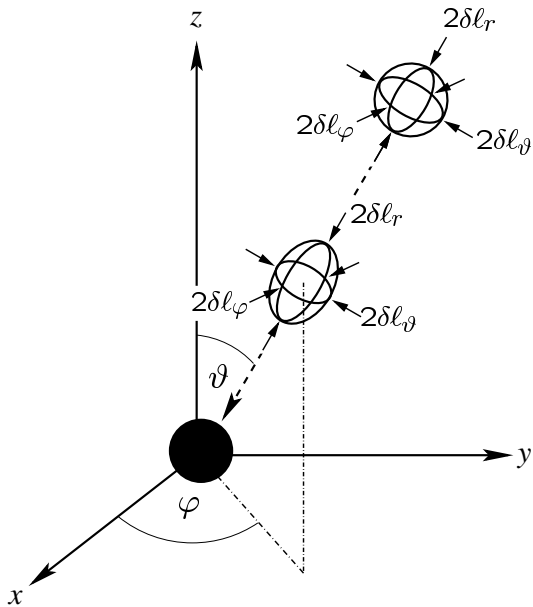}
\caption{Dust ball falling radially towards a spherically symmetric
  mass distribution (black sphere). Initially, the ball is spherical, a little
  later it deforms into an ellipsoid, the principal axes of which are
  aligned with the radial ($r$), poloidal ($\vartheta$) and azimuthal
  ($\varphi$) directions. For the purpose of representation,
  $\vartheta$ and $\varphi$ were chosen different from $\pi/2$ and 0,
  respectively, the values used in the text. Obviously, this is
  legitimate due to the spherical symmetry.  $\delta\ell_r$,
  $\delta\ell_\vartheta$, and $\delta\ell_\varphi$ are the lengths of
  the semiaxes of the (small) ellipsoid.}
\label{fig:radial_fall}
\end{figure}

\begin{figure}[t]
\includegraphics[width=8.5cm]{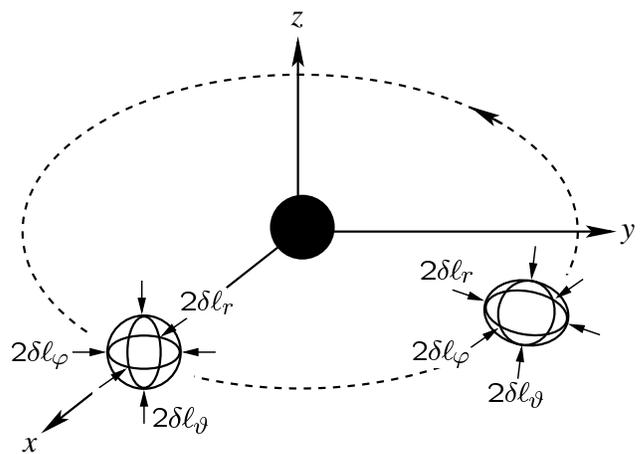}
\caption{Dust ball falling along a circular orbit about a spherical
  mass distribution. Again, the ball is spherical initially
  (configuration near the $x$ axis) and turns into an ellipsoid a very
  short time after. $\delta\ell\varphi$ remains however unchanged. Of course, the distance traveled while the ball
  keeps ellipsoidal shape may be much shorter than almost a quarter of
  the circle. This distance is exaggerated in the picture to allow
  well-separated drawings of the dust ball at two different times.
  The notation for lengths of semiaxes is the same as in
  Fig.~\ref{fig:radial_fall}. }
\label{fig:circular_fall}
\end{figure}

With the two highly symmetric configurations indicated, a dust ball
that falls radially as shown in Fig.~\ref{fig:radial_fall} and one
that is in a circular orbit, Fig.~\ref{fig:circular_fall}, the effort can
be even reduced further.

For a radially falling DB with initial
velocity zero, the direction of motion of all the falling particles is
towards the center of symmetry. Therefore, we need to know only by how
much the central particle and the particle at the end of a radial
semiaxis fall during a small time interval. The length changes of the
two semiaxes in the angular directions follow simply from the radial
interval fallen by the center particle multiplied by the angular
separation between that particle and the particle terminating the
considered semiaxis, because this angle remains unchanged during
radial fall of the particles. -- In the case of a circular orbit of a
DB, we know that the semiaxis along the direction of motion will not
change its length, so we have to calculate only the change of two
semiaxes.

Note that we cannot conclude from the DB vacuum equation that the
volume $V$ remains constant at \emph{all} times. At first sight, it
might be suggestive that if the volume of our ball does not
change for an infinitesimal time interval, we may infer its constancy
for arbitrary finite times. The reason this does not work is the side
condition that the test particles in the ball must be at rest with
respect to each other, initially. After the
first deformation of the ball the particles are no longer at rest
with respect to~each other. They are moving apart or closing in, so the
argument cannot be extended.

Before embarking on the calculation, let us do a practice example.
That is, we try to calculate the terms of the right-hand side of
Eq.~\eqref{eq:vol_change} in a case where we know what the result has
to be. This may reveal technical subtleties that have to be heeded.

\subsection{Dust ball falling freely in the Rindler metric}
\label{sec:db_rindler}
In this subsection, we will develop the correct procedure of
calculating the rate of change of a dust ball's diameter \emph{in the
  frame of its center particle}, considering such a ball in
an accelerating frame of reference.

The Rindler metric describes a set of accelerated observers, each of
which has a constant (in time) proper acceleration, while the
whole ensemble performs Born rigid motion. This means that observers
at different positions along the direction parallel to the motion must
experience different proper accelerations, so that their distance
shrinks, from the point of view of an inertial observer, just in the
right amount required by Lorentz contraction.  Then, in the comoving
frame of each Rindler observer distances between them remain constant.
The inertio-gravitational field\footnote{I use this notion that is
  close in spirit to Einstein's original ideas\protect\cite{janssen05}
  about gravity, because a number of contemporary authors would object
  to calling the field experienced by Rindler observers a
  gravitational one -- the spacetime of the Rindler metric is flat.  }
given by the Rindler metric is the closest analogy to a uniform
gravitational field that is possible in general relativity. Since the
metric can be obtained from the Minkowski metric by a global
coordinate transformation, it corresponds to flat spacetime and is
within the realm of SR.

This implies that we immediately know Eq.~\eqref{eq:vol_const} to be
satisfied in the Rindler metric for a freely falling dust
ball,\footnote{The metric is a vacuum solution to Einstein's
  equation.}  because the local inertial system in which its center
particle is at rest is even global, being described by the Minkowski
metric in all of spacetime. Hence, if the other dust particles are at
rest with respect to it at some initial time (and hence at rest
with respect to~each other), they will stay at rest with respect to~it forever. In fact,
this is even true for a dust ball of arbitrary size.  Nevertheless, we will
assume it to be small, because for a large ball the con\-dition of all
particles being at rest with respect to the cen\-ter particle will not
translate to being at rest in terms of Rindler coordinates; this 
only holds for objects in a suf\-fi\-cient\-ly local system. And of
course, the purpose of the practice calculation is to see whether we
can get the result known to be true on physical grounds using the
untransformed coordinates of the Rindler metric.

The line element reads
\begin{align}
  \D s^2 &= -\frac{\gR^2 x^2} {c^2} \D t^2 + \, \D x^2 + \D y^2 + \D
  z^2\>,
  \label{eq:rindler}
\end{align}
hence the Lagrangian is
\begin{align}
  -\frac{c^2}{2}= \frac12\left[-\frac{\gR^2 x^2} {c^2} \dot t^2 + \, \dot x^2 +
  \dot y^2 + \dot z^2\right]\>,
\end{align}
leading to the equations of motion for freely falling particles
\begin{align}
&\abl{}{\tau} \left(x^2 \dot t\right) = 0\>,
\label{eq:energy_intR}\\
&\ddot x =  -\frac{\gR^2 x} {c^2} \dot t^2\>,\qquad 
\ddot y = 0\>,\qquad \ddot z = 0\>.
\end{align}
Let the center particle satisfy the initial conditions $x(0)= x_0$,
$y(0)=z(0)=0$ and all particles from the ball $\dot x(0) = \dot y(0) =
\dot z(0) = 0$, then we obviously have for all particles
$y(\tau)=\text{const.}$ and $z(\tau)=\text{const.}$
For the center particle, we find $x^2 \dot t = \varepsilon =
\text{const.}$ and inserting this into the definition of the
Lagrangian, we get
\begin{align}
 -\frac{\gR^2 \varepsilon^2} {c^2 x^2}  + \dot x^2 = - c^2\>.
\label{eq:energy_cons_centR}
\end{align}
This determines $\varepsilon=c^2 x_0/\gR$. Next, we need an equation of
motion for a particle located initially at $x=x_0+\delta x(0)$. Note
that because this satisfies the initial condition $\delta \dot
x\rvert_{\tau=0} = 0$, its constant of motion (energy integral)
following from \eqref{eq:energy_intR} will not be the same value
$\varepsilon$ as for the center particle, so we cannot evaluate
$\delta x(\tau)$ by direct application of
\eqref{eq:energy_cons_centR}. However, multiplying through with $x^2$
and removing the constant of integration by taking the time
derivative, we obtain
\begin{align}
\ddot x = - \frac{c^2}{x} - \frac{\dot x^2}{x}\>,
\label{eq:xR_secnd_ord}
\end{align}
an equation that holds for all particles from the DB. The motion of
$\delta x$ is described by the so-called variational equation,
obtainable from \eqref{eq:xR_secnd_ord} by replacing $x$ with
$x+\delta x$, $\dot x$ with $\dot x+\delta \dot x$, etc. and
subtracting the equation for $x$ from that for $x+\delta x$. This is
the pedestrian's way. A route that is a bit faster is to use $\delta
f(x,\dot x, \ddot x) = \pabl{f}{x} \delta x + \pabl{f}{\dot x} \delta
\dot x +\pabl{f}{\ddot x}\delta \ddot x$, based on the fact that the
proper time $\tau$ is not varied along with the dependent variables.
We then find 
\begin{align}
  & \delta \ddot x = \left(\frac{c^2}{x^2}+\frac{\dot x^2}{x^2}
  \right)\delta x - \frac{2 \dot x}{x}\delta\dot x\>,
  \nonumber\\
  &\left.\frac{\delta \ddot x}{\delta x}\right\rvert_{\tau=0} =
  \frac{c^2}{x^2} >0 \>,
\end{align}
with the second line following from the fact that $\dot x(0)=0$. Since
the $y$ and $z$ coordinates of our particles remain constant, we
have $\delta \ddot y=\delta \ddot z=0$. Hence, if we calculate the rate of 
volume change according to \eqref{eq:vol_change} (replacing $\ell_a$
with $\delta x$, $\ell_b$ with $\delta y$, etc.), we will not get
zero!

Clearly, we must have missed something. The point to be observed here
is that \emph{after} a short time interval $\tau$, the center particle
will not be at rest, i.e., coordinate stationary in Rindler
coordinates anymore, it will have acquired a small velocity $-v=\ddot
x(0) \,\tau=-{c^2 \tau}/{x(0)}$. Correspondingly, the Rindler frame
accelerates with respect to~the freely falling frame, reaching velocity
$v$ at time $\tau$.

To obtain the spatial interval $\delta x_c$ in the frame of the center
particle, we have to apply a local Lorentz
trans\-formation\cite{cook04}
\begin{align}
  \delta x_c &= \gamma \,(\delta \tilde x +v \delta \tilde t)\>,
\quad
\delta t_c = \gamma \,(\delta\tilde  t + \frac{v}{c^2} \delta \tilde x)\>,
\end{align}
where $\delta\tilde x = \delta x$ and $\delta \tilde t = \frac{\gR
  x}{c}\, \delta t$ are the proper space and time intervals\footnote{At
  the beginning of the dust ball's free fall, $\tilde t$ may be
  identified with $\tau$ -- that is why it is correct to calculate the
  velocity as $\ddot x(0) \tau$. As soon as the dust ball center is
  not coordinate stationary anymore, $\tilde t$ and $\tau$ become
  different.}  of a fiducial Rindler observer and $\gamma =
1/\sqrt{1-\frac{v^2}{c^2}}$. $\delta t_c$ is the variation of the time
interval in the frame of the center particle.  But this is its proper
time and we consider variations at constant proper time, hence $\delta
t_c=\delta \tau=0$. Then we have $\delta \tilde t = -(v/c^2) \delta x$ and
\begin{align}
  \delta x_c &= \gamma \delta x \left(1-\frac{v^2}{c^2}\right) = \sqrt{1-\frac{v^2}{c^2}} \delta x \>.
\end{align}
Thus the spatial interval in the frame of the center particle is found 
 multiplying the  Rindler frame interval by an appropriate
Lorentz factor. The formula turns out to be the same as that for standard length
contraction.\footnote{Nevertheless, this is \emph{not} standard length
  contraction. The semiaxis of our dust ball in its direction of
  motion is \emph{maximum} in its rest frame, i.e., 
  in the frame of the center particle. In fact,
  $\delta x$ is not the length of \emph{any} object, it is the spatial
  interval between two events (or dust particles) at \emph{different}
  Rindler times, the time interval between them being $\delta t$.
  $\delta x_c$ is the corresponding interval at a \emph{fixed} proper
  time in the center particle frame and may therefore be interpreted
  as an extension of the dust ball at that time.  }
We can make $v$ as small as
we like by considering arbitrarily small time intervals $\tau$. But we
must take a second derivative which renders the effect
non-negligible. 
Expanding
\begin{align}
  \delta x_c = 
  \left(1-\frac{v^2}{2c^2}\right) \delta x = \left(1-\frac{c^2}{2x^2}
    \tau^2 \right) \delta x\>,
\end{align}
we obtain the second derivative as $\delta \ddot x - c^2/x^2\, \delta
x\rvert_{\tau=0}$ for small $\tau$, and this becomes exact for $\tau\to
0$, whence
\begin{align}
  &\left.\frac{\delta \ddot x_c}{\delta x_c}\right\rvert_{\tau=0} =
  \left.\frac{\delta \ddot x}{\delta x}\right\rvert_{\tau=0}
  -\left.\frac{c^2}{x^2}\right\rvert_{\tau=0} = 0\>,
\end{align}
which looks right. Plugging the expressions for ${\delta
  \ddot x_c}/{\delta x_c}$, ${\delta \ddot y_c}/{\delta y_c}$ and
${\delta \ddot z_c}/{\delta z_c}$, the latter two being unchanged
between the center particle frame and the Rindler frame, into 
formula \eqref{eq:vol_change} for the rate of volume change, we find zero, as
we must.

What our practice calculation reveals is that we have to take into
account velocity changes (referring to the coordinate stationary frame) of the
center particle between the beginning and the end of the small proper
time interval considered.  Equipped with this cautionary cue, we now
proceed to the central calculation.

\subsection{Distance rates of  change}
\label{sec:distchange}
In this subsection, we will obtain expressions for the rates of
change of the semiaxes of our dust balls, when they are aligned with
the axes of the spherical coordinate system.

We assume that the motion of the center particle remains in the
plane $\vartheta=\pi/2$, which simplifies equations a lot.  That
this is possible follows from the fact that
$\dot\vartheta=\ddot\vartheta =0$ satify Eq.~\eqref{eq:thetmotion}, if
$\vartheta=\pi/2$.

A spherical dust ball in radial free fall will elongate in the radial
direction and shrink in the two angular ones, due to the convergence
of the particle trajectories towards the center. Figure
\ref{fig:radial_fall} tries to convey this qualitatively.

For a dust ball orbiting the center of symmetry of the system ($r=0$)
on a circular trajectory, tidal forces will also tend to elongate the
ball in the radial direction, because particles farther from that
center will be too fast to be kept at the radius of the orbit, so the
will move outwards, and particles closer to the center will be too
slow for their orbit, so they are drawn inwards.  Particles ahead of
or behind, the center particle of the DB \emph{on} its equilibrium
trajectory will neither move away nor get closer to it.  Finally,
particles above and below the equatorial plane will be drawn towards
it, so there is shrinkage along the $\vartheta$ direction.  An attempt
was made to depict this behavior approximately in Fig.~\ref{fig:circular_fall}.

In both configurations, we may assume the semiaxes of the ellipsoid to be
oriented along the coordinate directions for symmetry reasons. If
the radial diameter of the ball is $2\delta r$ and its angular
extensions are $2\delta\vartheta$ and $2\delta\varphi$, respectively,
the semiaxes of the ellipsoid are given by
\begin{align}
\delta\ell_r &= \sqrt{h(r)} \delta r \>,
\\
\delta\ell_\vartheta &= r\delta \vartheta \>,
\\
\delta\ell_\varphi &= r\sin\vartheta\delta \varphi \>.
\end{align}
Initial conditions may be obtained from 
\begin{align}
\delta\dot\ell_r &= \frac{h'}{2\sqrt{h}} \dot r \delta r +  \sqrt{h} \delta \dot r \>,
\label{eq:dotellr}
\\
\delta\dot\ell_\vartheta &= \dot r\delta \vartheta + r \delta\dot\vartheta \>,
\label{eq:dotellthet}
\\
\delta\dot\ell_\varphi &=  \dot r\delta \varphi + r \delta\dot\varphi\>,
\label{eq:dotellphi}
\end{align}
where for brevity we drop the argument of the functions of $r$ and
where in the last equation, we already have used that $\vartheta
=\pi/2$.  If the particles of the ball are initially at rest with
respect to the center particle, $\delta\dot\ell_r$,
$\delta\dot\ell_\vartheta$, and $\delta\dot\ell_\varphi$ must all be
zero at $\tau=0$. Therefore, if $\dot r\rvert_{\tau=0}=0$, we obtain
as initial condition for the variation of $r$ $\ \delta\dot
r\rvert_{\tau=0}=0$, from Eq.~\eqref{eq:dotellr}.  This is true for
both configurations that we consider. Equally, we may conclude
$\delta\dot \vartheta\rvert_{\tau=0}=0$ and $\delta\dot
\varphi\rvert_{\tau=0}=0$ for particles aligned along the $\vartheta$
and $\varphi$ semiaxes.\footnote{But see the discussion of initial
  conditions for $\delta\dot\varphi$ in
  Sec.~\protect\ref{sec:circ_orbit}.} The treatment of, say, a DB
falling inward on a radial trajectory with an initially \emph{nonzero}
radial velocity would be much more complicated, because then the
initial first-order time derivatives of the variations of the three
variables would also be nonzero.

Because we need to compute second derivatives at $\tau=0$ only, we
immediately drop the vanishing first-order derivative terms in the
following expressions
\begin{align}
\frac{\delta\ddot\ell_r }{\delta\ell_r} &=\frac{h'}{2h} \ddot r + \frac{\delta \ddot r}{\delta r}\>,
\label{eq:ddotellr}
\\
\frac{\delta\ddot\ell_\vartheta }{\delta\ell_\vartheta} &=\frac{\ddot r}{r} 
+ \frac{\delta \ddot\vartheta}{\delta\vartheta}\>,
\label{eq:ddotellthet}
\\
\frac{\delta\ddot\ell_\varphi }{\delta\ell_\varphi} &=\frac{\ddot r}{r} + \frac{\delta \ddot\varphi}{\delta\varphi}\>.
\label{eq:ddotellphi}
\end{align}
To evaluate these formulas, we have to obtain $\ddot r$ from the
equations of motion for the center particle and $\delta\ddot r$,
$\delta\ddot \vartheta$, and $\delta\ddot\varphi$ from the equations of
motion for a particle displaced in one of the coordinate directions
with respect to~the center particle by either $\delta r$, $\delta
\vartheta$ or $\delta\varphi$. Since these displacements are small, we
may obtain the needed equations by linearization about the trajectory of the
center particle.

The sum of the three terms given by Eqs.~\eqref{eq:ddotellr} through
\eqref{eq:ddotellphi}, taken at $\tau=0$, may not yet be what we need as
the rate of volume change appearing in Eq.~\eqref{eq:vol_const}, because the
result refers to the coordinate stationary frame of our  metric. If the velocity of the center particle of the dust
ball changes during the small time interval considered, we have to
apply a correction using a Lorentz factor such as the one necessary in
the Rindler metric.

For the DB in a \emph{circular} orbit, neither the radial coordinate nor the
polar angle $\vartheta$ of the center particle change by free fall.
The azimuthal angle $\varphi$ changes in time, but its coordinate
velocity remains constant. We need not calculate any correction for
Eq.~\eqref{eq:ddotellphi} anyway, because we know by symmetry that
${\delta\ddot\ell_\varphi }/{\delta\ell_\varphi}=0$. The volume
change is calculable using Eqs.~\eqref{eq:ddotellr} and
\eqref{eq:ddotellthet} only.

In the case of the DB falling along a \emph{radial} trajectory, the
$\vartheta$ and $\varphi$ coordinates of the center particle remain
unchanged, so we do not need any correction either. Moreover, we know
that $\delta\ddot\vartheta=\delta\ddot\varphi=0$ in that case, because
all particles (starting with initial velocity zero) will fall towards
the center and hence, their angular coordinates will remain unchanged.
Tidal effects in the $\vartheta$ and $\varphi$ directions are trivial here.

However, we have to calculate a correction for the $r$ direction.
Multiplying $\delta\ell_r$ by the appropriate Lorentz factor
$\sqrt{1-\frac{v^2}{c^2}}=1-\frac12 \frac{h(r) \ddot r^2
  \tau^2}{c^2}$, we obtain for the relative rate of length change in the frame
of the center particle
\begin{align}
  \frac{\delta \ddot\ell_{cr}}{\delta \ell_{cr}} = \frac{\delta
    \ddot\ell_{r}}{\delta \ell_{r}} - \frac{h\ddot r^2}{c^2} =
  \frac{\delta \ddot r}{\delta r}+\frac{h'}{2h} \ddot r - \frac{h
    \ddot r^2}{c^2}\>,
\label{eq:ddotellrc}
\end{align}
everything to be evaluated at $\tau=0$.

\subsection{Radial fall of a dust ball}
\label{sec:rad_fall_DB}
In this subsection, we will find a constraint on the metric by imposing
Eq.~\eqref{eq:vol_const} on the volume rate of change of a dust ball
falling radially, as in Fig.~\ref{fig:radial_fall}, from an initial
``rest'' state. 

We put the center of our ball of test particles at the initial radius
$r_0$, take $\vartheta=\pi/2$ and may take, without restriction of
generality, $\varphi=0$ as initial azimuthal angle. Then the ball is
dropped with zero initial velocity with respect to a local coordinate
stationary observer. This defines the initial rest state.

Equations \eqref{eq:thetmotion} and \eqref{eq:phimotion} give us
$\dot\vartheta=\dot\varphi=0$. To completely determine the motion we
consider Eqs.~\eqref {eq:tmotion} and \eqref{eq:spherical_lagrangian},
which provide
\begin{align}
\dot t &= \frac{\varepsilon_r}{f(r)}\>,  \qquad \varepsilon_r=\text{const.}
\\
 \dot r^2 &= \frac{c^2}{h(r)} \left(\frac{\varepsilon_r^2}{f(r)}-1\right)\>,
\label{eq:dotr_rad}
\end{align}
and the condition $\dot r=0$ at $r=r_0$ determines the constant:
$\varepsilon_r=\sqrt{f(r_0)}$. Since the equation contains only a
single constant of motion, we may obtain a second-order equation of
motion for all dust particles involving $r$ alone simply by isolating
the constant [multiplying Eq.~\eqref{eq:dotr_rad} by $f(r) h(r)$] and
taking the proper time derivative:
\begin{align}
\ddot r = - c^2 \frac{f'}{2fh} - \frac{(fh)'}{2fh} \dot r^2\>.
\label{eq:ddotr_rad}
\end{align}
From this emerges one of the quantities required in the evaluation of \eqref{eq:ddotellr}:
\begin{align}
\ddot r(0) = \left.- c^2 \frac{f'}{2fh}\right\rvert_{r=r_0}\>.
\end{align}
Taking the variation of \eqref{eq:ddotr_rad}, we get
\begin{align}
\delta\ddot r  = -c^2 \left(\frac{f'}{2fh}\right)'\delta r - \left(\frac{(fh)'}{2fh} \right)' \dot r^2 \delta r - 2\frac{(fh)'}{2fh}\dot r \delta \dot r\>.
\end{align}
We need this only at $\tau=0$, where $\dot r=0$, so we drop the second and third terms, which leads to
\begin{align}
\delta\ddot r\rvert_{\tau=0} = -\frac{c^2}{2h}\left[\frac{f''}{f} - \frac{f'}{f}\left(\frac{f'}{f}+\frac{h'}{h}\right)\right] \delta r\>.
\end{align}
Since particles from the ends of the $\vartheta$ and $\varphi$
semiaxes of our ellipsoid fall at constant angle $\delta\vartheta$
and $\delta\varphi$, respectively, we infer from \eqref{eq:ddotellthet}
and \eqref{eq:ddotellphi}
\begin{align}
  \frac{\delta\ddot\ell_\vartheta }{\delta\ell_\vartheta} =
  \frac{\delta\ddot\ell_\varphi }{\delta\ell_\varphi} = \frac{\ddot
    r}{r}\>.
\end{align}
Putting everything together, we find
\begin{align}
  \left.\frac{\ddot{V}}{V}\right\rvert_{\tau=0} &= \left.\frac{\delta
      \ddot\ell_{cr}}{\delta \ell_{cr}}\right\rvert_{0} +
  \left.\frac{\delta\ddot\ell_\vartheta }{\delta\ell_\vartheta}
  \right\rvert_{0}+ \left.\frac{\delta\ddot\ell_\varphi
    }{\delta\ell_\varphi}\right\rvert_{0}
  \nonumber\\
  &= \left.-\frac{c^2}{2h}\left[\frac{f''}{f} -
      \frac{f'}{f}\left(\frac{f'}{f}+\frac{h'}{h}\right)\right] +
    \frac{h'}{2h} \ddot r - h \frac{\ddot r^2}{c^2} + \frac{2 \ddot
      r}{r}\rule{0mm}{7mm}\>\right\rvert_{0}
  \nonumber\\
  &= -\frac{c^2}{2h}\frac{f''}{f} +\frac{c^2}{4h}
  \frac{f'}{f}\left(\frac{f'}{f}+\frac{h'}{h}\right) - c^2
  \frac{f'}{fh r} \stackrel{!}{\,=} 0\>.
\label{eq:firstdiffeq}
\end{align}
Of course, the value of $r$ is $r_0$ in the last equation, but since
$r_0$ can be taken arbitrarily, we may drop the subscript.  The last
line of Eq.~\eqref{eq:firstdiffeq} is a differential equation in $r$
that the metric functions must satisfy in order for the DB vacuum equation \eqref{eq:vol_const} to hold. It can be
simplified a bit (multiplying by $-2 f h/c^2 f'$), which yields
\begin{align}
\frac{f''}{f'} -\frac12 \left(\frac{f'}{f}+\frac{h'}{h}\right) + \frac2r = 0\>.
\label{eq:first}
\end{align}
In principle, this completes the analysis of a dust ball falling
radially. However, we realize that our differential equation is second
order for $f$, so we will need at least two boundary conditions for
$f$ -- and one for $h$ -- when finally trying to solve it.  One
boundary condition for each of the two functions is trivial: we require the line element
\eqref{eq:spherical_static} to become Min\-kows\-kian as $r\to\infty$.
Then we have $\lim_{r\to\infty} f(r)=1$ and $\lim_{r\to\infty}
h(r)=1$. 

A second boundary condition for $f$ may be obtained as follows:
Eq.~\eqref{eq:dotr_rad} has the form of a one-dimensional law of
energy conservation in Newtonian mechanics. Setting $r_0=\infty$
($\Rightarrow \varepsilon_r=1$) and multiplying by $m/2$, we get
\begin{align}
\frac{m}{2} \dot r^2 - \frac{m c^2}{2h(r)}\left(\frac{1}{f(r)} -1\right)=0\>,
\end{align}
which we require to become identical, for sufficiently large $r$, to
the corresponding law of energy conservation obtained from Newton's
law of gravitation (for a particle of mass $m$ the velocity of which
becomes zero at infinity):
\begin{align}
\frac{m}{2} \dot r^2 - G \frac{m M}{r} = 0\>.
\end{align}
$G$ is Newton's gravitational constant.  Using $h(r)\sim 1$ $ (r\to\infty)$, we have
${1}/{f(r)} -1 \sim {2 G M}/{c^2 r}$ $(r\to\infty)$ and
\begin{align}
f(r) &\sim \frac1{1+ \frac{2 G M}{c^2 r}} \sim 1-  \frac{2 G M}{c^2 r}\quad  (r\to\infty)\>.
\label{eq:asymptotics_f}
\end{align}
This second boundary condition that $f(r)$ must satisfy at
large $r$  introduces the Schwarzschild radius $r_S = {2 G
  M}/{c^2}$.

\subsection{Dust ball in circular orbit}
\label{sec:circ_orbit}
In this subsection, we repeat the procedure of
Sec.~\ref{sec:rad_fall_DB} for a dust ball circling the mass
distribution, as in Fig.~\ref{fig:circular_fall}, with all dust grains
initially having the same velocity as the center particle (a notion
that makes sense because of the closeness of the particles). A second
constraint on the metric will be obtained.

The motion of a particle in a circular orbit $r=r_0$ is characterized by two
constants of motion, arising via integration of Eqs.~\eqref{eq:tmotion}
and \eqref{eq:phimotion}
\begin{align}
\dot t &= \frac{\varepsilon_c}{f(r_0)}\>,\qquad \varepsilon_c=\text{const.}
\\
r_0^2 \dot\varphi &= n = \text{const.}
\end{align}
They describe conservation of energy and of angular momentum,
respectively. The first equation implies $\dot t= \text{const.}$, the
second $ \dot\varphi=\omega = \text{const.}$ $\omega$ is the angular
frequency of the particle referred to its proper
time.\footnote{The angular frequency as seen by a coordinate
  stationary observer at $r_0$ will be smaller.}

It is then convenient to use these constants in the de\-finit\-ion of the
Lagrangian \eqref{eq:spherical_lagrangian} and in the radial geodesic
equation \eqref{eq:rmotion} to determine their values in terms of the
radius $r_0$ of the orbit (using $\dot r= \ddot r= 0$):
\begin{align}
&-c^2 \frac{\varepsilon_c^2}{f(r_0)} +r_0^2 \omega^2 = -c^2 \>,\nonumber \\
 &\frac{f'(r_0)}{2} c^2 \frac{\varepsilon_c^2}{f(r_0)^2} - r_0 \omega^2 =0 \>,
\end{align}
which yields (we drop radial arguments again)
\begin{align}
\varepsilon_c^2 &= \frac{f^2}{f-r_0 f'/2}\>,\qquad
\omega^2 = \frac{c^2}{2 r_0} \frac{f'}{f-r_0 f'/2}\>.\label{eq:def_eps_c_omega_c}
\end{align}
This completes the description of the trajectory of the dust ball's
center particle.

As to the variations giving the trajectories of slightly displaced
particles, we start with the equation that is simplest to treat. This
is Eq.~\eqref{eq:thetmotion}. Because $\dot\vartheta = 0$ and
$\vartheta=\pi/2$, terms of the equation in which neither $\vartheta$
nor its derivative are varied must vanish.  Thus we have
\begin{align}
  &\abl{}{\tau}\left[r^2 \delta\dot\vartheta\right] - r^2  \dot\varphi^2
  \underbrace{\abl{}{\vartheta}\left(\sin\vartheta\cos\vartheta\right)
  }_{\cos 2\vartheta = -1}\delta \vartheta = 0\>,
  \nonumber\\
 & \delta\ddot\vartheta + \omega^2 \delta \vartheta = 0\>.
\end{align}
This equation can be integrated exactly and shows particles that
are slightly displaced in the direction of the polar angle to
oscillate about the equatorial plane, as long as they remain close
enough to the center particle.  We are interested only in the short-time behavior.
From Eq.~\eqref{eq:ddotellthet} and the fact that $\ddot r = \ddot r_0
= 0$, we note
\begin{align}
  \frac{\delta\ddot\ell_\vartheta }{\delta\ell_\vartheta} =
  \frac{\delta\ddot\vartheta}{\delta\vartheta} = -\omega^2\>.
\label{eq:ddot_theta_c}
\end{align}
To arrive at the corresponding result for the radial direction is a bit
more tedious than in the case of purely radial fall, because the
variational equations for all variables but $\vartheta$ are coupled.
Taking the variations of Eqs.~\eqref{eq:rmotion},
\eqref{eq:spherical_lagrangian}, and \eqref{eq:phimotion} and dropping
all terms containing a factor $\dot r=0$, $\dot\vartheta=0$,
$\cos\vartheta=0$ or $\ddot\varphi=0$, we get
\begin{align}
  \delta\ddot r &= -\frac12 \left(\frac{f'}{h}\right)' c^2 \dot t^2 
  \delta r - \frac{f'}{h} c^2 \dot t \delta \dot t + 
  \left(\frac{r}{h}\right) '\dot \varphi^2 \delta r + \frac{2 
    r}{h}\dot \varphi \delta\dot\varphi\>,
\label{eq:var_ddotr_circ}
  \\
0 &= -f' c^2 \dot t^2 \delta r - 2 f c^2 \dot t\delta\dot t + 2r \dot \varphi^2 \delta r +
 2 r^2 \dot \varphi\delta\dot \varphi \>,
\label{eq:var_lagr_circ}
\\
\delta \ddot\varphi &= -\frac2r  \dot \varphi\delta \dot r \>.
\label{eq:var_ddotphi_circ}
\end{align}
\emph{After} having varied the coordinates, we may use the integrals
of motion for the variables referring to the center particle (and we
set $r=r_0$).  Eq.~\eqref{eq:var_lagr_circ} then simplifies to
\begin{align}
\varepsilon_c c^2 \delta\dot t = r_0^2 \omega\delta\dot\varphi\>,
\end{align}
i.e., the $\delta r$ term vanishes. We may use this equation to
remove $\delta\dot t$ from \eqref{eq:var_ddotr_circ}, which then
reduces to
\begin{align}
  \delta\ddot r &= \delta r \left[-\frac12
    \left(\frac{f''}{h}-\frac{f'h'}{h^2}\right) c^2
    \frac{\varepsilon_c^2}{f^2} + \left(\frac1h-\frac{r_0
        h'}{h^2}\right) \omega^2\right]
  \nonumber \\
  &+\delta\dot\varphi \left[-\frac{f'}{fh} r_0^2 \omega + \frac{2
      r_0}{h}\omega\right]\>.
\label{eq:deltaddotr_c1}
\end{align}
We would like to eliminate the factor $\delta\dot\varphi$ from this
equation, too.  It looks as if this might be achieved via integration of
Eq.~\eqref{eq:var_ddotphi_circ}, but to do so, we need an initial
condition for $\delta\dot\varphi$. Note that this cannot be taken from
Eq.~\eqref{eq:dotellphi}, which tells us that $\delta\dot\varphi=0$
\emph{along} the $\varphi$ direction.  The initial conditions
discussed in Sec.~\ref{sec:distchange} referred to particles displaced
from the center particle along a local coordinate axis, i.e., the
other two coordinates were kept constant.
However, here we need an initial condition for
$\delta\dot\varphi$ as $r$ changes.
Fortunately, the situation is simple enough to infer the correct
condition easily. The velocity component along the $\varphi$
direction of a selected particle relative to the center particle is
given, as long as the relative motion is slow, by $r \dot\varphi -
r_0\omega$.  A condition for initial relative rest is then $r
\dot\varphi\rvert_{\tau=0}= r_0\omega$ or, if expressed in terms of
the displacement coordinates $\delta r$ and $\delta \varphi$
\begin{align}
\left.\delta r  \dot\varphi\right\rvert_{\tau=0} +
 \left.r \delta\dot\varphi\right\rvert_{\tau=0} =0\>.
\end{align}
Since we have to solve \eqref{eq:deltaddotr_c1} only for an
infinitesimal time interval, this initial condition may be directly
used to express $\delta\dot\varphi$ in the equation. It is not
necessary to first solve \eqref{eq:var_ddotphi_circ}. Essentially, all
the coefficient functions of the differential equation
\eqref{eq:deltaddotr_c1} are only required at $\tau=0$. Setting
$\delta\dot\varphi = - \delta r \,\omega/r_0$, we obtain
\begin{align}
  \frac{\delta\ddot r}{\delta r} &= -\frac12
    \left(\frac{f''}{h}-\frac{f'h'}{h^2}\right) c^2
    \frac{\varepsilon_c^2}{f^2} + \omega^2\left(\frac1h-\frac{r_0
        h'}{h^2} +\frac{r_0 f'}{fh}-\frac{2}{h}\right)
\>.
\label{eq:deltaddotr_c2}
\end{align}
Finally, we insert the definitions \eqref{eq:def_eps_c_omega_c}  of $\varepsilon_c^2$ and $\omega^2$ into Eq.~\eqref{eq:deltaddotr_c2} to find
\begin{align}
  \frac{\delta\ddot\ell_r }{\delta\ell_r} &= \frac{\delta\ddot r}{\delta r} =
\frac{c^2}{2h(f-r_0
    f'/2)}\left(-f''-\frac{f'}{r_0}+\frac{f'^2}{f}
 \right)\>,
\end{align}
wherefrom, after combination with \eqref{eq:ddot_theta_c}, we arrive at
\begin{align}
  \left.\frac{\ddot{V}}{V}\right\rvert_{\tau=0} &=
  \left.\frac{\delta\ddot\ell_r }{\delta\ell_r}\right\rvert_{0} +
  \left.\frac{\delta\ddot\ell_\vartheta
    }{\delta\ell_\vartheta}\right\rvert_{0}
  \nonumber\\
  &= \frac{c^2}{2(f-r_0
    f'/2)}\left[\frac1{h}\left(-f''-\frac{f'}{r_0}+\frac{f'^2}{f}\right)
    -\frac{f'}{r_0} \right]
  \nonumber\\
  &\stackrel{!}{=} 0\>.
\end{align}
Again, we may simplify a bit by removing common prefactors, and we
rename $r_0$ to $r$. Our second equation for the
functions $f$ and $h$ then reads
\begin{align}
- \frac{h}{r} &= \frac{f''}{f'} - \frac{f'}{f} + \frac1r\>.
\label{eq:second}
\end{align}

\section{Solution for the metric}
\label{sec:solution_metric}

All that remains to be done is to solve the system of ordinary
differential equations \eqref{eq:first} and \eqref{eq:second}. The
fact that in both equations the explicit appearance of the independent
variable is in the inverse form $1/r$ suggests that a simplification
may arise, if we transform to the new variable $u=1/r$.
Setting
\begin{align}
f(r) &= F(u)\>, & h(r) &= H(u) \>,
\nonumber\\
\Rightarrow \>\> f'(r) &= -u^2 F'(u) \>, & h'(r) &= -u^2 H'(u) \>,
\nonumber\\
f''(r) &= 2u^3  F'(u) + u^4 F''(u) \>, 
\end{align}
the equations become
\begin{align}
\frac{F''}{F'} &=\frac12\left(\frac{F'}{F}+\frac{H'}{H}\right)\>,
\label{eq:first_tr} 
\\
H &= 1+ u \left(\frac{F''}{F'}-\frac{F'}{F}\right)\>,
\label{eq:second_tr} 
\end{align}
where we have again suppressed the argument for brevity.
Equation~\eqref{eq:first_tr} can be directly integrated once:
\begin{align}
\ln F' &= \frac12 \ln FH + \tilde \alpha \quad\Rightarrow\quad
F'^2 = \alpha^2 FH\>.
\label{eq:FH_prod}
\end{align}
Herein, $\tilde\alpha$ and $\alpha$ are constants of integration
($\alpha=\EXP{\tilde\alpha}$). Inserting $H$ from
\eqref{eq:second_tr}, we obtain an equation for $F$ alone:
\begin{align}
F'^2 &=\alpha^2 \left[F+ u\left(\frac{F'' F}{F'}-F'\right)\right]\>.
\label{eq:diffeq_F}
\end{align}
To find the \emph{general} solution to this equation is
difficult.\footnote{It is, however, not impossible. On presentation of
  the equation to the computer algebra system MAPLE 17, the latter
  spat out a general solution dependent on two constants of
  integration, in an implicit form. Obtaining an explicit form would
  require the solution of a transcendental equation to invert a
  function.} On the other hand, a particular solution may be found by
inspection:  obviously,  there must be a solution of the
form $F'=\text{const.}$; the $F''$ term vanishes in this case, and the
product of $u$ and $-F'$ just cancels the linear term of $F$. The
remainder of the equation is a relationship between constants.
Let us then set $F=A u + B$ and plug it into Eq.~\eqref{eq:diffeq_F}. This gives
\begin{align}
A^2 = \alpha^2 B \quad \Rightarrow \quad F(u) &= \alpha \sqrt{B} u +B
\nonumber\\
\Rightarrow \quad f(r) &=  \frac{\alpha \sqrt{B}}{r} +B
\end{align} 
and using the asymptotics \eqref{eq:asymptotics_f}, we can read off
both constants ($B=1$, $\alpha=-r_S$). As it turns out, the solution
obtained by inspection is precisely the one we need, it satisfies the
physical boundary conditions.

Therefore, $F(u) = 1- r_S u$, and from \eqref{eq:FH_prod}, we find
\begin{align}
H(u) = \frac{F'^2}{\alpha^2 F} = \frac{1}{1- r_S u}\>.
\end{align}
We end up with both metric functions, and hence the Schwarzschild
metric, being determined:
\begin{align}
  f(r)& = 1- \frac{r_S}{r}\>, \qquad h(r) = \frac{1}{1-r_S/r}\>.
\end{align}

\section{Conclusions}
\label{sec:conclusions}

What we have accomplished is a derivation of the Schwarzschild metric
in a way I would consider completely \emph{physics first}. The
fundamental law Eq.~\eqref{eq:vol_const}, on which the derivation is
based, does not have the form of a field equation. It is expressed in
terms of physical objects, balls of test particles that can be easily
visualized. 

The derivation is shorter than calculations based on the traditional
tensor calculus, such as Schwarzschild's original
one\cite{schwarzschild16a}. When antisymmetry is built into the tensor
formalism as in the theory of differential forms, the resulting gain
in efficiency permits derivations that are more concise. In Gr{\o}n and
Hervik's book\cite{gron07}, the calculation is done in four pages --
but not before page 215. A lot of mathematics has to be learned up to
that point, whereas the approach given here uses standard calculus.

Its main advantage lies, however, in its transparency and visual
appeal. What happens physically can be easily imagined. The only
purely mathematical step is the final solution of the two differential
equations in Sec.~\ref{sec:solution_metric}.

There are two ways to present the approach. In the first, no reference
to the field equations is needed at all.  The DB vacuum equation may
be taken as the postulate of a new physical law. Indeed, this
postulate is easy to motivate, because it also holds in Newtonian
physics, with proper time replaced by the absolute Newtonian time and
without restriction to a particularly moving frame -- volumes are
frame independent in Newtonian physics. Given this law, the \emph{derivation}
of Newton's universal law of gravitation is a three-liner, assuming
spherical symmetry of the potential to be determined.

Once it is accepted as the physical meaning underlying Newton's law of
gravitation outside a mass distribution, its generalization to the
relativistic case is straightforward and follows a standard scheme
when going from classical mechanics to SR: replace
time by proper time and specify the (inertial) frame, in which the law
holds. This gives the form \eqref{eq:vol_const} of the DB vacuum
equation.

To motivate the \emph{full} DB law of gravitation starting from a Newtonian
version is less straightforward. It is clear that the mass density
should be replaced by something involving energy density and it is
plausible due to the special relativistic relationships in which
energy is only part of a four vector that momentum flow must enter
the equations as well, which leads to the appearance of
pressure.\cite{baez05} However, it takes a leap of faith to be sure
that energy and pressure (or stresses) will appear in the DB law
precisely as stated. On the other hand, if we wish to just derive the
Schwarzschild metric, we need only the DB vacuum equation,
 for the validity of which simple arguments are available. If
cosmological problems are to be treated by the method as well, as is
done in Ref.~\onlinecite{baez05}, then the full DB law is needed and,
instead of making it a postulate, it may be preferable to present it
as a lookahead to Einstein's equation.

The postulational approach would be particularly useful in an undergraduate course in
which the field equations were to be omitted completely.

Alternatively, if the field equations are presented in the course
anyway, which will certainly be true in most graduate courses, a
second way of presentation may be more appropriate. Spell out the DB
law of gravitation, state it to be a particular formulation of the
physical law that will be expressed in terms of partial differential
equations later, promise a rigorous derivation of the DB law then, and
proceed with physical applications.  The Schwarzschild metric will
thus not appear out of the blue but find some justification from an
underlying law. Cosmological models may be discussed without first
introducing the Riemann curvature tensor.

It might be added that a difference in philosophy between the DB law
and the Einstein equation is that the former is a Lagrangian
description, working in the frame of the dust ball, whereas the latter
is Eulerian in nature. Lagrangian descriptions tend to be simpler
locally, but their extension to all of space is not as natural as that
of Eulerian ones. In complicated cases, global solutions will be more
easily obtained within a Eulerian framework. The Schwarzschild case is
simple enough to be solved in a Lagrangian scheme as well, here even
in one that gets by without partial differential equations.

\rule{0pt}{0pt}

\newcommand{\phre}[1]{Phys. Rev. E {\bf #1}}
\newcommand{\phrl}[1]{Phys. Rev. Lett. {\bf #1}}


\begin{thebibliography}{24}%
\makeatletter
\providecommand \@ifxundefined [1]{%
 \@ifx{#1\undefined}
}%
\providecommand \@ifnum [1]{%
 \ifnum #1\expandafter \@firstoftwo
 \else \expandafter \@secondoftwo
 \fi
}%
\providecommand \@ifx [1]{%
 \ifx #1\expandafter \@firstoftwo
 \else \expandafter \@secondoftwo
 \fi
}%
\providecommand \natexlab [1]{#1}%
\providecommand \enquote  [1]{``#1''}%
\providecommand \bibnamefont  [1]{#1}%
\providecommand \bibfnamefont [1]{#1}%
\providecommand \citenamefont [1]{#1}%
\providecommand \href@noop [0]{\@secondoftwo}%
\providecommand \href [0]{\begingroup \@sanitize@url \@href}%
\providecommand \@href[1]{\@@startlink{#1}\@@href}%
\providecommand \@@href[1]{\endgroup#1\@@endlink}%
\providecommand \@sanitize@url [0]{\catcode `\\12\catcode `\$12\catcode
  `\&12\catcode `\#12\catcode `\^12\catcode `\_12\catcode `\%12\relax}%
\providecommand \@@startlink[1]{}%
\providecommand \@@endlink[0]{}%
\providecommand \url  [0]{\begingroup\@sanitize@url \@url }%
\providecommand \@url [1]{\endgroup\@href {#1}{\urlprefix }}%
\providecommand \urlprefix  [0]{URL }%
\providecommand \Eprint [0]{\href }%
\providecommand \doibase [0]{http://dx.doi.org/}%
\providecommand \selectlanguage [0]{\@gobble}%
\providecommand \bibinfo  [0]{\@secondoftwo}%
\providecommand \bibfield  [0]{\@secondoftwo}%
\providecommand \translation [1]{[#1]}%
\providecommand \BibitemOpen [0]{}%
\providecommand \bibitemStop [0]{}%
\providecommand \bibitemNoStop [0]{.\EOS\space}%
\providecommand \EOS [0]{\spacefactor3000\relax}%
\providecommand \BibitemShut  [1]{\csname bibitem#1\endcsname}%
\let\auto@bib@innerbib\@empty
\bibitem [{\citenamefont {Hartle}(2003)}]{hartle03}%
  \BibitemOpen
  \bibfield  {author} {\bibinfo {author} {\bibfnamefont {J.~B.}\ \bibnamefont
  {Hartle}},\ }\href@noop {} {\emph {\bibinfo {title} {{Gravity. An
  Introduction to Einstein's General Relativity}}}}\ (\bibinfo  {publisher}
  {Addison-Wesley},\ \bibinfo {address} {San Francisco},\ \bibinfo {year}
  {2003})\BibitemShut {NoStop}%
\bibitem [{\citenamefont {Chen}(2005)}]{chen05}%
  \BibitemOpen
  \bibfield  {author} {\bibinfo {author} {\bibfnamefont {T.}~\bibnamefont
  {Chen}},\ }\href@noop {} {\emph {\bibinfo {title} {{Relativity, Gravitation
  and Cosmology}}}}\ (\bibinfo  {publisher} {Oxford University Press},\
  \bibinfo {address} {Oxford},\ \bibinfo {year} {2005})\BibitemShut {NoStop}%
\bibitem [{\citenamefont {Hartle}(2006)}]{hartle06}%
  \BibitemOpen
  \bibfield  {author} {\bibinfo {author} {\bibfnamefont {J.~B.}\ \bibnamefont
  {Hartle}},\ }\bibfield  {title} {\enquote {\bibinfo {title} {{General
  relativity in the undergraduate physics curriculum}},}\ }\href@noop {}
  {\bibfield  {journal} {\bibinfo  {journal} {Am. J. Phys.}\ }\textbf {\bibinfo
  {volume} {74}},\ \bibinfo {pages} {14--21} (\bibinfo {year}
  {2006})}\BibitemShut {NoStop}%
\bibitem [{Note1()}]{Note1}%
  \BibitemOpen
  \bibinfo {note} {I will use the terms \protect \emph {Einstein's field
  equations}, \protect \emph {Einstein's equation}, and just \protect \emph
  {field equations} interchangeably. They all mean the same thing.}\BibitemShut
  {Stop}%
\bibitem [{\citenamefont {Christensen}\ and\ \citenamefont
  {Moore}(2012)}]{christensen12}%
  \BibitemOpen
  \bibfield  {author} {\bibinfo {author} {\bibfnamefont {N.}~\bibnamefont
  {Christensen}}\ and\ \bibinfo {author} {\bibfnamefont {T.}~\bibnamefont
  {Moore}},\ }\bibfield  {title} {\enquote {\bibinfo {title} {{Teaching general
  relativity to undergraduates}},}\ }\href@noop {} {\bibfield  {journal}
  {\bibinfo  {journal} {Physics Today}\ }\textbf {\bibinfo {volume}
  {65\textmd(6)}},\ \bibinfo {pages} {41--47} (\bibinfo {year}
  {2012})}\BibitemShut {NoStop}%
\bibitem [{\citenamefont {Moore}(2013)}]{moore13}%
  \BibitemOpen
  \bibfield  {author} {\bibinfo {author} {\bibfnamefont {T.~A.}\ \bibnamefont
  {Moore}},\ }\href@noop {} {\emph {\bibinfo {title} {{A General Relativity
  Workbook}}}}\ (\bibinfo  {publisher} {University Science Books},\ \bibinfo
  {address} {Sausalito},\ \bibinfo {year} {2013})\BibitemShut {NoStop}%
\bibitem [{\citenamefont {Schild}(1960)}]{schild60}%
  \BibitemOpen
  \bibfield  {author} {\bibinfo {author} {\bibfnamefont {A.}~\bibnamefont
  {Schild}},\ }\bibfield  {title} {\enquote {\bibinfo {title} {{Equivalence
  Principle and Red-Shift Measurements}},}\ }\href@noop {} {\bibfield
  {journal} {\bibinfo  {journal} {Am. J. Phys.}\ }\textbf {\bibinfo {volume}
  {28}},\ \bibinfo {pages} {778--780} (\bibinfo {year} {1960})}\BibitemShut
  {NoStop}%
\bibitem [{\citenamefont {Gruber}\ \emph {et~al.}(1988)\citenamefont {Gruber},
  \citenamefont {Price}, \citenamefont {Matthews}, \citenamefont {Cordwell},\
  and\ \citenamefont {Wagner}}]{gruber88}%
  \BibitemOpen
  \bibfield  {author} {\bibinfo {author} {\bibfnamefont {R.~P.}\ \bibnamefont
  {Gruber}}, \bibinfo {author} {\bibfnamefont {R.~H.}\ \bibnamefont {Price}},
  \bibinfo {author} {\bibfnamefont {S.~M.}\ \bibnamefont {Matthews}}, \bibinfo
  {author} {\bibfnamefont {W.~R.}\ \bibnamefont {Cordwell}}, \ and\ \bibinfo
  {author} {\bibfnamefont {L.~F.}\ \bibnamefont {Wagner}},\ }\bibfield  {title}
  {\enquote {\bibinfo {title} {{The impossibility of a simple derivation of the
  Schwarzschild metric}},}\ }\href@noop {} {\bibfield  {journal} {\bibinfo
  {journal} {Am. J. Phys.}\ }\textbf {\bibinfo {volume} {56}},\ \bibinfo
  {pages} {265--269} (\bibinfo {year} {1988})}\BibitemShut {NoStop}%
\bibitem [{\citenamefont {Kassner}(2015)}]{kassner15}%
  \BibitemOpen
  \bibfield  {author} {\bibinfo {author} {\bibfnamefont {K.}~\bibnamefont
  {Kassner}},\ }\bibfield  {title} {\enquote {\bibinfo {title} {{Classroom
  reconstruction of the Schwarz\-schild metric}},}\ }\href@noop {} {\bibfield
  {journal} {\bibinfo  {journal} {Eur. J. Phys.}\ }\textbf {\bibinfo {volume}
  {36}},\ \bibinfo {pages} {065031 (1--20),} (\bibinfo {year}
  {2015})}\BibitemShut {NoStop}%
\bibitem [{\citenamefont {Kassner}(2016)}]{kassner16}%
  \BibitemOpen
  \bibfield  {author} {\bibinfo {author} {\bibfnamefont {K.}~\bibnamefont
  {Kassner}},\ }\href@noop {} {\enquote {\bibinfo {title} {{How to obtain the
  Schwarzschild metric before Einstein's field equations}},}\ } (\bibinfo
  {year} {2016}),\ \Eprint {http://arxiv.org/abs/arXiv:1602.08309v2 [gr-qc]}
  {arXiv:1602.08309v2 [gr-qc]} \BibitemShut {NoStop}%
\bibitem [{\citenamefont {Baez}\ and\ \citenamefont {Bunn}(2005)}]{baez05}%
  \BibitemOpen
  \bibfield  {author} {\bibinfo {author} {\bibfnamefont {J.~C.}\ \bibnamefont
  {Baez}}\ and\ \bibinfo {author} {\bibfnamefont {E.~F.}\ \bibnamefont
  {Bunn}},\ }\bibfield  {title} {\enquote {\bibinfo {title} {{The meaning of
  Einstein's equation}},}\ }\href@noop {} {\bibfield  {journal} {\bibinfo
  {journal} {Am. J. Phys.}\ }\textbf {\bibinfo {volume} {73}},\ \bibinfo
  {pages} {644--652} (\bibinfo {year} {2005})}\BibitemShut {NoStop}%
\bibitem [{Note2()}]{Note2}%
  \BibitemOpen
  \bibinfo {note} {Note that in general, the condition of the dust particles in
  the ball initially being at rest with respect to each other makes sense only
  for a sufficiently small ball. General relativity does not allow us to assign
  a physical meaning to relative velocities, hence a relative state of rest,
  for objects that are not very close to each other.}\BibitemShut {Stop}%
\bibitem [{Note3()}]{Note3}%
  \BibitemOpen
  \bibinfo {note} {In standard relativistic Lagrangian mechanics, the action of
  a free particle between two events is, up to a prefactor, the integral of the
  line element $\DOTSI \intop \ilimits@ _1^2 \protect \mathrm {d}s$. The
  Lagrangian is then proportional to $\protect \mathrm {d}s/\protect \mathrm
  {d}t$. The Lagrangian \protect \textup {\hbox {\mathsurround \z@ \protect
  \normalfont (\ignorespaces \ref {eq:spherical_lagrangian}\unskip
  \@@italiccorr )}} is, apart from a prefactor, the square of this, but with
  the arbitrary time coordinate $t$ replaced by the proper time $\tau$. It can
  be shown that, if an affine parameter $\tau $, provi\-ded by the proper time
  for massive particles, is chosen as time coordinate in the action integral,
  extremalisation of $\DOTSI \intop \ilimits@ _1^2 (\protect \mathrm
  {d}s/\protect \mathrm {d}\tau )^2 \protect \mathrm {d}\tau $ yields the same
  equations of motion as extremalisation of $\DOTSI \intop \ilimits@ _1^2
  \protect \mathrm {d}s/\protect \mathrm {d}t\protect \tmspace +\thinmuskip
  {.1667em}\protect \mathrm {d}t$. Therefore, our Lagrangian produces the
  correct equations of motion -- and it is easier to use than the standard
  Lagrangian, avoiding the appearance of certain square roots. The factor
  $\protect \frac {1}{2}$ has been introduced for convenience, to cancel out
  some factors of 2, appearing in taking derivatives. Finally, that $L$ is
  constant is of course due to the fact that $L=\protect \frac 12 (\protect
  \mathrm {d}s/\protect \mathrm {d}\tau )^2$ and that $\protect \mathrm {d}s^2
  = -c^2\protect \mathrm {d}\tau ^2$ for massive particles.}\BibitemShut
  {Stop}%
\bibitem [{Note4()}]{Note4}%
  \BibitemOpen
  \bibinfo {note} {I use this notion that is close in spirit to Einstein's
  original ideas\protect \cite {janssen05} about gravity, because a number of
  contemporary authors would object to calling the field experienced by Rindler
  observers a gravitational one -- the spacetime of the Rindler metric is
  flat.}\BibitemShut {Stop}%
\bibitem [{Note5()}]{Note5}%
  \BibitemOpen
  \bibinfo {note} {The metric is a vacuum solution to Einstein's
  equation.}\BibitemShut {Stop}%
\bibitem [{\citenamefont {Cook}(2004)}]{cook04}%
  \BibitemOpen
  \bibfield  {author} {\bibinfo {author} {\bibfnamefont {R.~J.}\ \bibnamefont
  {Cook}},\ }\bibfield  {title} {\enquote {\bibinfo {title} {{Physical time and
  physical space in general relativity}},}\ }\href@noop {} {\bibfield
  {journal} {\bibinfo  {journal} {Am. J. Phys.}\ }\textbf {\bibinfo {volume}
  {72}},\ \bibinfo {pages} {214--219} (\bibinfo {year} {2004})}\BibitemShut
  {NoStop}%
\bibitem [{Note6()}]{Note6}%
  \BibitemOpen
  \bibinfo {note} {At the beginning of the dust ball's free fall, $\protect
  \mathaccentV {tilde}07Et$ may be identified with $\tau $ -- that is why it is
  correct to calculate the velocity as $\protect \mathaccentV {ddot}07Fx(0)
  \tau $. As soon as the dust ball center is not coordinate stationary anymore,
  $\protect \mathaccentV {tilde}07Et$ and $\tau $ become
  different.}\BibitemShut {Stop}%
\bibitem [{Note7()}]{Note7}%
  \BibitemOpen
  \bibinfo {note} {Nevertheless, this is \protect \emph {not} standard length
  contraction. The semiaxis of our dust ball in its direction of motion is
  \protect \emph {maximum} in its rest frame, i.e., in the frame of the center
  particle. In fact, $\delta x$ is not the length of \protect \emph {any}
  object, it is the spatial interval between two events (or dust particles) at
  \protect \emph {different} Rindler times, the time interval between them
  being $\delta t$. $\delta x_c$ is the corresponding interval at a \protect
  \emph {fixed} proper time in the center particle frame and may therefore be
  interpreted as an extension of the dust ball at that time.}\BibitemShut
  {Stop}%
\bibitem [{Note8()}]{Note8}%
  \BibitemOpen
  \bibinfo {note} {But see the discussion of initial conditions for $\delta
  \protect \mathaccentV {dot}05F\varphi $ in Sec.~\protect \ref
  {sec:circ_orbit}.}\BibitemShut {Stop}%
\bibitem [{Note9()}]{Note9}%
  \BibitemOpen
  \bibinfo {note} {The angular frequency as seen by a coordinate stationary
  observer at $r_0$ will be smaller.}\BibitemShut {Stop}%
\bibitem [{Note10()}]{Note10}%
  \BibitemOpen
  \bibinfo {note} {It is, however, not impossible. On presentation of the
  equation to the computer algebra system MAPLE 17, the latter spat out a
  general solution dependent on two constants of integration, in an implicit
  form. Obtaining an explicit form would require the solution of a
  transcendental equation to invert a function.}\BibitemShut {Stop}%
\bibitem [{\citenamefont {Schwarzschild}(1916)}]{schwarzschild16a}%
  \BibitemOpen
  \bibfield  {author} {\bibinfo {author} {\bibfnamefont {K.}~\bibnamefont
  {Schwarzschild}},\ }\bibfield  {title} {\enquote {\bibinfo {title} {{\"Uber
  das Gravitationsfeld eines Massenpunktes nach der Einsteinschen Theorie}},}\
  }in\ \href@noop {} {\emph {\bibinfo {booktitle} {{Sitzungsberichte der
  K\"oniglich-Preu{\ss}ischen Akademie der Wissenschaften}}}}\ (\bibinfo
  {publisher} {Reimer},\ \bibinfo {address} {Berlin},\ \bibinfo {year} {1916})\
  pp.\ \bibinfo {pages} {189--196},\ \bibinfo {note} {{English translation:
  \emph{On the Gravitational Field of a Mass Point According to Einstein's
  Theory}, S. Antoci and A. Loinger, arXiv:physics/9905030v1}}\BibitemShut
  {NoStop}%
\bibitem [{\citenamefont {Gr{\o}n}\ and\ \citenamefont
  {Hervik}(2007)}]{gron07}%
  \BibitemOpen
  \bibfield  {author} {\bibinfo {author} {\bibfnamefont {\O.}\ \bibnamefont
  {Gr{\o}n}}\ and\ \bibinfo {author} {\bibfnamefont {S.}~\bibnamefont
  {Hervik}},\ }\href@noop {} {\emph {\bibinfo {title} {{Einstein's General
  Theory of Relativity: With Modern Applications in Cosmology}}}}\ (\bibinfo
  {publisher} {Springer Science \& Business Media},\ \bibinfo {address}
  {Springer, Berlin},\ \bibinfo {year} {2007})\BibitemShut {NoStop}%
\bibitem [{\citenamefont {Janssen}(2005)}]{janssen05}%
  \BibitemOpen
  \bibfield  {author} {\bibinfo {author} {\bibfnamefont {M.}~\bibnamefont
  {Janssen}},\ }\bibfield  {title} {\enquote {\bibinfo {title} {{Of pots and
  holes: Einstein's bumpy road to general relativity}},}\ }\href@noop {}
  {\bibfield  {journal} {\bibinfo  {journal} {Ann. Phys. (Leipzig)}\ }\textbf
  {\bibinfo {volume} {14}},\ \bibinfo {pages} {58--85} (\bibinfo {year}
  {2005})}\BibitemShut {NoStop}%
\end{thebibliography}
\end{document}